\documentclass[8.5pt,twoside,twocolumn]{article}
\usepackage[super,sort&compress,comma]{natbib}
\usepackage{amsmath,amssymb}
\usepackage{times,mathptm}
\usepackage{sectsty}
\usepackage{balance}
\usepackage{color}

\usepackage[dvips]{graphicx}
\usepackage{lastpage}
\usepackage[format=plain,justification=raggedright,singlelinecheck=false,font=small,labelfont=bf,labelsep=space]{caption}
\usepackage{fancyhdr}
\begin{document}

\thispagestyle{plain}
\fancypagestyle{plain}{
%\fancyhead[L]{\includegraphics[height=8pt]{headers/LH.eps}}
%\fancyhead[C]{\hspace{-1cm}\includegraphics[height=20pt]{headers/CH.eps}}
%\fancyhead[R]{\includegraphics[height=10pt]{headers/RH.eps}}
\renewcommand{\headrulewidth}{1pt}}
\renewcommand{\thefootnote}{\fnsymbol{footnote}}
\renewcommand\footnoterule{\vspace*{1pt}%
\hrule width 3.4in height 0.4pt \vspace*{5pt}}

\makeatletter
\renewcommand\@biblabel[1]{#1}
\renewcommand\@makefntext[1]%
{\noindent\makebox[0pt][r]{\@thefnmark\,}#1}
\makeatother
\renewcommand{\figurename}{\small{Fig.}~}
\sectionfont{\large}
\subsectionfont{\normalsize}

\fancyfoot{}
%\fancyfoot[LO,RE]{\vspace{-6pt}\includegraphics[height=8.5pt]{headers/LF.eps}}
%\fancyfoot[CO]{\vspace{-6.5pt}\hspace{11.4cm}\includegraphics{headers/RF.eps}}
%\fancyfoot[CE]{\vspace{-6.6pt}\hspace{-12.7cm}\includegraphics{headers/RF.eps}}
\fancyfoot[RO]{\footnotesize{\sffamily{1--\pageref{LastPage} ~\textbar  \hspace{2pt}\thepage}}}
\fancyfoot[LE]{\footnotesize{\sffamily{\thepage~\textbar\hspace{4.4cm} 1--\pageref{LastPage}}}}
\fancyhead{}
\renewcommand{\headrulewidth}{1pt}
\renewcommand{\footrulewidth}{1pt}
\setlength{\arrayrulewidth}{1pt}
\setlength{\columnsep}{6.5mm}
\setlength\bibsep{1pt}

\twocolumn[
  \begin{@twocolumnfalse}
\noindent\LARGE{\textbf{Interaction of Spherical Colloidal Particles in Nematic Media with Degenerate Planar Anchoring}}
\vspace{0.6cm}

\noindent\large{\textbf{Mohammad Reza Mozaffari\textit{$^{a}$},
Mehrtash Babadi\textit{$\,^{b}$}, Jun-ichi Fukuda\textit{$\,^{c}$},
and Mohammad Reza
Ejtehadi$\,^{\ast}$\textit{$\,^{a}$}}}\vspace{0.5cm}
%Please note that \ast indicates the corresponding author(s) but no footnote text is required.

%\noindent\textit{\small{\textbf{Received Xth XXXXXXXXXX 20XX, Accepted Xth XXXXXXXXX 20XX\newline
%First published on the web Xth XXXXXXXXXX 200X}}}
%\noindent \textbf{\small{DOI: 10.1039/b000000x}}
%\vspace{0.6cm}
%Please do not change this text.

\noindent \normalsize{The interaction between two spherical colloidal particles
with degenerate planar anchoring in a nematic media is studied by numerically minimizing
the bulk Landau-de Gennes and surface energy using a finite element method. We find that the
energy achieves its global minimum when the particles are in
close contact and making an angle $\theta = 28^\circ \pm 2$ with respect to the bulk nematic director,
in agreement with the experiments. Although the quadrupolar structure of the director
field is preserved in the majority of configurations, we show that for smaller orientation angles
and at smaller inter-particle separations, the axial symmetry of the topological defect-pairs
is continuously broken, resulting in the emergence of an attractive interaction}.

\vspace{0.5cm}
 \end{@twocolumnfalse}
  ]
\footnotetext{\textit{$^{a}$~Department of Physics, Sharif University of Technology, P.O. Box 14588-89694 Tehran, Iran.}}
\footnotetext{\textit{$^{b}$~Department of Physics, Harvard University, Cambridge, MA 02138, USA.}}
\footnotetext{\textit{$^{c}$~Nanosystem Research Institute, National Institute of Advanced Industrial Science and Technology (AIST),1-1-1 Umezono, Tsukuba 305-8568, Japan.}}
\footnotetext{\textit{$^\ast$ ejtehadi@sharif.edu}}

\section{Introduction}

Studying the behavior of colloidal particles in anisotropic fluids
with long-range orientational ordering, such as nematic liquid
crystals, has attracted a great attention in soft condensed matter
physics~\cite{Poulin.science.275.97,Nazarenko.PhysRevLett.87.01,
Makoto.PhysRevLett.92.04,Smalyukh.PhysRevLett.95.05,Musevic.science.313.06,Fukuda.JPhysSoc.Japan.78.09}.
The orientation order parameter of the fluid (e.g. director of
nematic liquid crystal) is distorted from its uniform orientation in
the bulk due to anchoring to the surface of the colloidal particles.
These elastic distortions create topological defects around the
particles~\cite{Poulin.PhysRevE.57.98} and induce anisotropic long
and short range interactions between the
particles~\cite{Ruhwandl.PhysRevE.55.97,Lubensky.PhysRevE.57.98}.

Depending on the colloidal material and its coating, the surrounding
fluid may have a normal orientation (\textit{normal or homeotropic
anchoring}), or parallel orientation (\textit{planar anchoring})
with respect to the colloidal surfaces. For normal anchoring, the
orientation of the fluid is locally and uniquely determined on the
colloidal surfaces. For planar anchoring, the orientation of the
fluid is degenerate on the colloidal surfaces and is determined by
the global structure of the fluid. However, the director field in
the surface of the particles is affected by environment for any
finite anchoring, such a freedom makes the theoretical
investigations more complicated for planar anchoring.

In case of a single colloidal particle, the particle-defect pair
induces a dipolar or a quadrupolar long-range elastic distortion
field~\cite{stark.PhysRep.351.01,lev.PhysRevE.65.02} depending on
the anchoring type. The long-range dipolar structure results from a
\textit{satellite} point defect, when the size of the particle is
large compared to the coherence length of the nematic fluid and the
anchoring is
normal~\cite{Lubensky.PhysRevE.57.98,Poulin.PhysRevE.57.98}. The
quadrupolar configuration appears in both normal and planar
anchorings. In the normal case, a disclination ring  (\textit{saturn
ring}) encircles the particle, when the size of the particle is
small or the strength of anchoring is
weak~\cite{Ruhwandl.PhysRevE.55.97,Poulin.PhysRevE.57.98}. In planar
anchoring, the elastic distortions form two point defects
(\textit{boojums}) at the poles of the particles, aligned along the
nematic
direction~\cite{Ruhwandl.PhysRevE.55.97,Poulin.PhysRevE.57.98}.

The more physically interesting configurations are achieved when
there are many colloidal particles present in the medium. For large
separations of particles, the defects around of each of the
particles is independent of that of the other particles, and
(anisotropic) interaction potential between them is determined by
the long-range orientational field of the fluid. In this regime, and
in case of two particles separated by a distance $d$, the effective
interaction potential between them is proportional to $d^{-3}$ or
$d^{-5}$ for dipolar or quadrupolar defect configurations,
respectively~\cite{Poulin.science.275.97,Ruhwandl.PhysRevE.55.97,Poulin.PhysRevE.57.98,Makoto.PhysRevLett.92.04}.
When the particles approach each other, the defect structures are
distorted and the interactions deviate from the far-field
dipolar-dipolar or quadrupolar-quadrupolar
interactions~\cite{Guzman.PhysRevLett.91.03,Andrienko.PhysRevE.68.03,Silvestre.JPhysCondMat.16.04,Ravnik.PhysRevLett.99.07,Fernandez-Nieves.PhysRevLett.99.07,
Fukuda.PhysRevLett.94.05,Takahashi.PhysRevE.77.08}.

Experimentally, the colloidal interactions in nematic liquid crystals are studied
using optical~\cite{Smalyukh.PhysRevLett.95.05} or magneto-optical
tweezers~\cite{kotar.PhysRevLett.96.06, vilfan.PhysRevLett.101.08}.
The medium-induced interactions play an essential role in the formation of
chain~\cite{Smalyukh.PhysRevLett.95.05} or
crystal~\cite{Nazarenko.PhysRevLett.87.01,Nych.PhysRevLett.98.07,Smalyukh.PhysRevLett.93.04}
suspensions of the colloids (or droplets) in nematic liquid crystals.

Smalyukh \textit{et al.}~\cite{Smalyukh.PhysRevLett.95.05} have measured the
angular and the radial components of the force between two
particles with planar anchoring in a nematic liquid crystal as a function
of the inter-particle separation $d$, the angle between the bulk nematic
director and the vector connecting the particles, $\theta$ (see Fig.~\ref{geo1}).
They observe deviation from the far-field theoretical
quadrupole-quadrupole interaction~\cite{Ruhwandl.PhysRevE.55.97}
when the objects are in the close-contact regime. They also find
that the equilibrium configuration corresponds to a very small separation
of particles, $d \simeq 2R$, and $\theta \simeq 30^\circ$.

In this paper, we study the fluid-induced interaction between two
spherical colloidal particles of radius $R$ by numerically
minimizing the sum of elastic Landau-de Gennes free energy of the
bulk fluid~\cite{deGennes.95} and the degenerate planar anchoring
surface energy introduced by Fournier {\em et
al}~\cite{Fournier.EurophysLett.72.05}. In particular, we are
interested in the regime of strong anchoring and large particles
($R$ large compared to the coherence length of the fluid, $\xi$).
Our main goal is to study the close-contact configurations for which
no theoretical work has been done to our knowledge, though
interesting physics is expected to emerge due to the strong
interaction of the topological defects. We also aim to explain the
experimentally observed angle of $\theta \simeq 30^\circ$ at
equilibrium.

The paper is organized as follows. The theoretical model is
explained in Section~\ref{model}. We describe the details of our
numerical approach in Section~\ref{numerical}. We finally present
and discuss the results in Section~\ref{results} and summarize our
findings in Section~\ref{conc}.

\section{The Model}\label{model}
\begin{figure}[t]
\begin{center}
\includegraphics[width=5.0cm, angle=0]{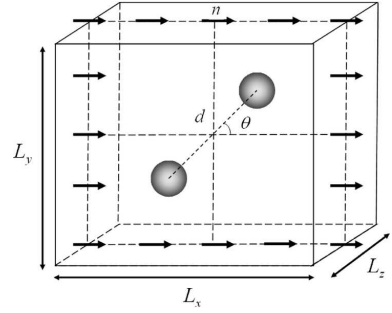}
\end{center}
\caption{Schematic representation of the studied system. The nematic
director is fixed in $x$ direction at the boundaries and two
particles are symmetrically placed in $x$-$y$ plane with respect to the
center of the box.} \label{geo1}
\end{figure}

The geometry of the studied system is schematically illustrated in
Fig.~\ref{geo1}. We consider two identical spherical colloidal
particles with radius $R$ immersed in a 3D nematic cell. The nematic
director is aligned along $x$-axis at the boundaries of the cell. We
scale all the lengths with respect to the radius of the particles.
The dimensions of the cell is $L_{x}=15R$, $L_{y}=15R$ and
$L_{z}=6R$. The centers of the particles are confined to the plane
$z=L_{z}/2$ and are separated by a center-to-center distance $d$.
The line joining the center of the particles makes an angle $\theta$
with $x$-axis. The dimensions of the cell are chosen in a way to
ensure that in all of the studied configurations, the distance
between the boundaries and the particles is much larger than the
coherence length and big enough so that the nematic director
distribution is not affected by the boundaries. The free energy of
the system can be written as:
\begin{equation}
F(d,\theta) = F_{\mathrm{b}}(d,\theta) + F_{\mathrm{s}}(d,\theta) + F_{\mathrm{c-c}}(d),
\end{equation}
where $F_{\mathrm{b}}(d,\theta)$ is bulk nematic fluid free energy , $F_{\mathrm{s}}(d,\theta)$
is the surface energy and $F_{\mathrm{c-c}}(d)$ is the Van der Waals colloid-colloid interaction.
The free energy functionals will be described in detail in the following sections.

In realistic situations, the only appreciable effect of
Van der Waals colloid-colloid interactions is to provide a short-range repulsion between the colloidal particles.
Such effects become relevant only in the regime where the separation of the colloid surfaces approaches
the atomic length-scales. In this study, we confine ourselves to the regime where the surface separations are
larger than the nematic coherence length, i.e. $d - 2R \gg \xi \gg 1 \mathrm{\AA}$. We note that the surface
separations can chosen to be appreciably smaller than the size of the particles in this regime. Therefore,
due to the separation of scales in this regime, we ignore
the Van der Waals interaction between the colloids in this study.

\subsection{The Nematic Order Parameter}
The nematic fluid is described by a local, $3\times3$, traceless and symmetric
tensor order parameter,
$Q_{ij}=S(\hat{n}_{i}\hat{n}_{j}-\delta_{ij}/3)$, which can be
specified by five independent components,
\begin{equation}
\label{tensor}
\mathbf{Q} =
\left(\begin{array}{ccc}
Q_{11}  &  Q_{12}  &  Q_{13} \\
Q_{12}  &  Q_{22}  &  Q_{23} \\
Q_{13}  &  Q_{23}  &  Q_{33}
\end{array} \right),
\end{equation}
where $Q_{33}=-(Q_{11}+Q_{22})$. The scalar order parameter, $S$,
and the director orientation, $\hat{n}$, are locally obtained by the
largest eigenvalue of the tensor order parameter, $\lambda_{\rm max}
= \frac{2}{3}S$, and its corresponding eigenvector, respectively.
We note that working with a tensor order parameter and expanding the free energy
in its terms allows the formation of biaxial order, which is a necessary
ingredient for a realistic description of topological defects and their interactions.

\subsection{The Bulk Free Energy}\label{bulk}
The bulk free energy of the nematic fluid is described well by the
Landau-de Gennes model~\cite{deGennes.95}, in which the free energy functional
is expanded in powers of the tensor order parameter and its spatial
derivatives:
\begin{equation}
\begin{split}
F_{\mathrm{b}} &=\int _{\Omega }\mathrm{d}V\, \bigg(\frac{A}{2}Q_{ij}Q_{ji}-\frac{B}{3}Q_{ij}Q_{jk}Q_{ki}+\frac{C}{4}\left(Q_{ij}Q_{ji}\right)^2\\ & \quad \quad \quad+\frac{L_{1}}{2}\partial_{k}Q_{ij}\partial_{k}Q_{ij} + \frac{L_{2}}{2}{\partial_{j}Q_{ij}\partial_{k}Q_{ik}} \bigg),
\end{split}
\end{equation}
where the indices refer to Cartesian coordinates, the summation over repeated indices
is assumed and $\Omega$ denotes the volume occupied by the
nematic liquid crystal. The first three terms are the Landau-de
Gennes free energy which describe the bulk isotropic-nematic (IN)
transition. The coefficients $A$, $B$, and $C$ are the
material-dependent parameters.

The derivative terms are the contribution of the elastic free energy
in the nematic phase. The nematic elastic constants, $L_1$ and
$L_2$, are related to the Frank elastic constants by
$L_{1}=K_{\mathrm{twist}}\big/2S^{2}$ and
$L_{2}=(K_{\mathrm{splay}}-K_{\mathrm{twist}})\big/2S^{2}=(K_{\mathrm{bend}}-K_{\mathrm{twist}})\big/2S^{2}$.
In this study we restrict ourselves to one-elastic constant
approximation that means all the Frank elastic constants should be
equivalent which leads to $L_{2}=0$.

To simplify calculations, we rescale the tensor order parameter as
$q=(4B/3\sqrt{6}C)^{-1}Q$, such that
$q_{ij}=\hat{S}(\hat{n}_{i}\hat{n}_{j}-\delta_{ij}/3)$, where
$\hat{S}=(4B/3\sqrt{6}C)^{-1} S$. As a
consequence, the dimensionless free energy becomes
\begin{equation}
\begin{split}
\hat{f}_{\mathrm{b}} =\frac{F_{\mathrm{b}}}{f_{0}R^3} &=\int_{\Omega}\mathrm{d}\hat{V}\,\bigg(\frac{\tau }{2}q_{ij}q_{ji}-\frac{\sqrt{6}}{4}q_{ij}q_{jk}q_{ki}+\frac{1}{4}\left(q_{ij}q_{ji}\right)^2\\ & \quad \quad \quad +\frac{1}{2}\hat{\xi}^{2}\hat{\partial}_{k}q_{ij}\hat{\partial}_{k}q_{ij} \bigg),
\end{split}
\label{dimentionless}
\end{equation}
where $f_0=C(4B/3\sqrt{6}C)^4$, $\tau=27\text{AC}/8B^2$ is effective
dimensionless temperature, $\xi=\sqrt{27L_{1}C/8B^2}$ is the nematic
coherence
length~\cite{Fukuda.PhysRevE.69.04,Fukuda.MolCrystLiq.Cryst.435.05}.
In these dimensionless units, the fluid undergoes a first-order
isotropic-nematic transition at $\tau=1/8$. The isotropic phase
becomes unstable for $\tau<0$ . The scalar nematic order parameter
in bulk is given by
$\hat{S}_{\mathrm{b}}=(3\sqrt{6}/16)\left(1+\sqrt{1-64\tau
/9}\right)$. Throughout this paper, the quantities appearing with a
hat are rescaled with respect to the the radius of the particles,
e.g. $\mathrm{d}\hat{V}=R^{-3}\mathrm{d}V$, $\hat{\xi} = R^{-1}\xi$
and $\hat{\partial}_{k}=R\partial_{k}$. Following the related
previous
studies~\cite{Fukuda.PhysRevE.69.04,Fukuda.MolCrystLiq.Cryst.435.05,Ravnik.PhysRevLett.99.07},
the dimensionless temperature and length scales were set to
$\tau=(3\sqrt{6}-8)/12$ and $\hat{\xi}=0.03$ respectively. This
choice of parameters closely match the parameters of the widely used
liquid crystal mesogen 5CB and results in formation of stable
topological
defects~\cite{Fukuda.PhysRevE.69.04,Fukuda.MolCrystLiq.Cryst.435.05}.

\subsection{The Surface Free Energy}
As mentioned in the introduction, the preferred anchoring can be
modeled with much more
ease~\cite{Rapini.JPhysFrance.30.54,Nobili.PhysRevA.46.92} compared
to the planar degenerate anchoring due to uniqueness of the
orientation of nematic fluid in the vicinity of the colloidal
surfaces.

Fournier \textit{et al.}~\cite{Fournier.EurophysLett.72.05} have recently introduced a
two-parameter surface energy functional that is bounded from below and assumes its
minimum in the manifold of degenerate planar configurations.
The surface energy consists of two terms, controlling the planar anchoring and
fixing the scalar order parameter on the surface. Since we expect the formation surface topological
(e.g. two surface defects of charge $+1/2$ in case of a single colloid), we relax the
second constraint like previous studies~\cite{vilfan.PhysRevLett.101.08},
resulting in the following single parameter surface energy functional:
\begin{equation}
\label{surface_free_energy}
F_{\mathrm{s}} = W\int _{\partial\Omega}\mathrm{d}A\,(\tilde{Q}_{ij}-\tilde{Q}^{\bot}_{ij})(\tilde{Q}_{ji}-\tilde{Q}^{\bot}_{ji}),
\end{equation}
where $\tilde{Q}_{ij}=Q_{ij} + S{\delta_{ij}/3}$, $\tilde{Q}^{\bot}_{ij}=(\delta_{ik}-\hat{\nu}_{i}\hat{\nu}_{k})\tilde{Q}_{kl}(\delta_{lj}-\hat{\nu}_{l}\hat{\nu}_{j})$ is the projection of $\tilde{Q}_{ij}$ onto the tangent
plane of the surface, and $\hat{\nu}$ is the normal to the surface. $W$
is positive and controls the stiffness of anchoring. The surface energy
can be written in dimensionless variables:
\begin{equation}
\begin{split}
\hat{f}_{\mathrm{s}} =\frac{F_{\mathrm{s}}}{f_{0}R^3} &=\frac{w}{R}\int_{\partial\Omega}\mathrm{d}\hat{A}\,(\tilde{q}_{ij}-\tilde{q}^{\bot}_{ij})(\tilde{q}_{ji}-\tilde{q}^{\bot}_{ji}),
\end{split}
\end{equation}
where $w=27WC/8B^{2}$, $\tilde{q}_{ij}=(4B/3\sqrt{6}C)^{-1}\tilde{Q}_{ij}$, $\tilde{q}^{\bot}_{ij}=(\delta_{ik}-\hat{\nu}_{i}\hat{\nu}_{k})\tilde{q}_{kl}(\delta_{lj}-\hat{\nu}_{l}\hat{\nu}_{j})$ and $\mathrm{d}\hat{A}=R^{-2}\mathrm{d}A$. We chose
$\hat{w}=0.0156$, which describes strong planar anchoring to the surface once we
take the bulk free energy parameters into account.

\begin{figure}
\begin{center}
\includegraphics[width=4.0cm, angle=0]{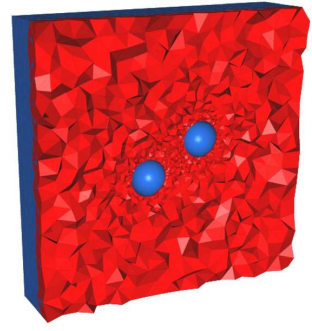}
\end{center}
\caption{A typical tetrahedral mesh used in the FEM analysis. The mesh is finer
near the colloidal surfaces to capture a more accurate representation
of the curved surfaces and to provide increased numerical accuracy near
the topological defects.} \label{geo2}
\end{figure}

\section{Numerical Minimization of the Free Energy}\label{numerical}
We adopt a finite element method (FEM) approach to minimize the free
energy functional described in the preceding sections. The nematic
cell was decomposed into tetrahedral elements by using the automatic
mesh generator~\emph{Gmsh}~\cite{gmsh.Online}. In order to capture a
more accurate representation of the curved surfaces and to provide
increased numerical accuracy near the topological defects which are
expected to be formed on or in the vicinity of the surfaces, a finer
mesh size of $L_{\mathrm{SMS}} = 0.025R$ was used near the shperical
boundaries. We note that since the nematic coherence length is $\xi
= 0.03R$, the physics of elastic deformations can be properly
captured in the used mesh. The mesh size was increased to
$L_{\mathrm{LMS}} = 0.25R$ away from the particles in order to
reduce the computational cost (see Fig.~\ref{geo2}). The tensor
order parameter was linearly interpolated within each of elements in
the evaluation of the free energy integrals. We note that linear
interpolation is the simplest scheme that preserves the properties
of $Q$ as an order tensor. For each configuration of the particles,
the total dimensionless free energy was minimized using a conjugate
gradient (CG) method~\cite{Press.92}, yielding the Effective
Potential Energy (EPE) $U(d,\theta)$. In order to accelerate the
minimization procedure for any given ($\theta$, $d$), we used the
relaxed director profile from the closest preceding configuration as
the initial guess. The configuration space of the particles was
scanned in the range $2.10R \leq d \leq 3.5R$ and $0^\circ \leq
\theta \leq 90^\circ$ in steps of $0.05R$ and $2^\circ$
respectively. For larger separations, $3.5R<d\leq6.0R$, we scan with
larger $\theta$ steps of $10^\circ$. For each configuration, the
minimization procedure was stopped when the relative free energy
improvements dropped below $10^{-5}$. We present and discuss the
results in the next section.

\section{Results and Discussion}\label{results}

\begin{figure}
\centering
\includegraphics[width=5.0cm, angle=-90]{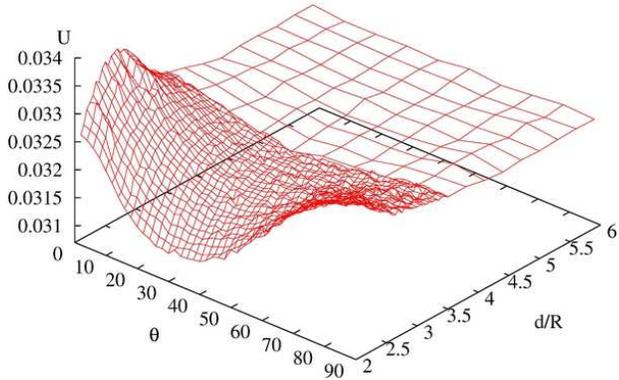}
\caption{The free energy landscape of the system
of two particles as a function of the inter-particles distance, $d/R$, and $\theta$, the
angle between the line joining the center of particles and the
far-field director.} \label{pot-3D}
\end{figure}

\begin{figure}
\centering
\includegraphics[width=6.0cm, angle=-90]{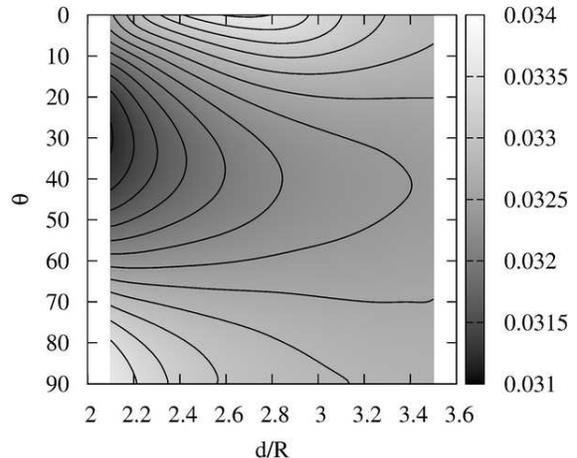}
\caption{The contour plot of the EPE between two
particles as a function of the inter-particles distance, $d/R$, and
$\theta$, the angle between the line joining the center of particles
and the far-field director.} \label{pot-contour}
\end{figure}

\begin{figure}
\begin{center}
\includegraphics[width=6.0cm, angle=-90]{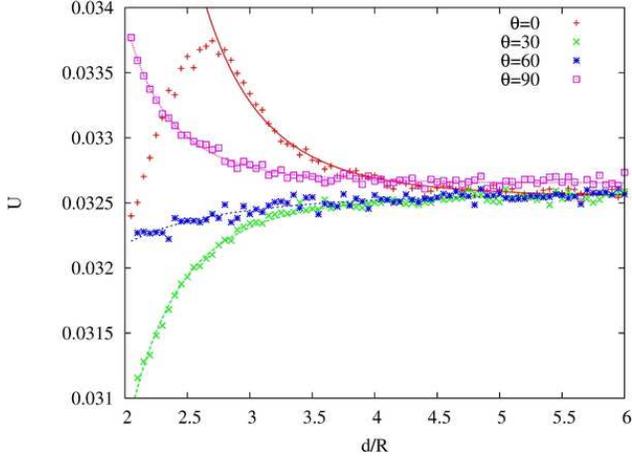}
\caption{Effective potential energy between two particles as a
function of the inter-particles distance, $d/R$, for four different
orientations $\theta= 0^\circ$, $30^\circ$, $60^\circ$ and
$90^\circ$.}\label{pot-distance}
\end{center}
\end{figure}

Figure~\ref{pot-3D} shows the full equilibrium free energy landscape of
the system. The smooth contour plot of the free energy landscape is shown
in Fig.\ref{pot-contour} for better clarity. The normal to the
contour lines specify the direction of the net force between the colloidal
particles. The effective potential energy of the system of two colloids is
also shown in Fig.~\ref{pot-distance} as a function of the particle
separation $d$, for four different orientations. The figure shows that for
$\theta=90^\circ$, the particles repel each other in the whole range
of inter-particle separations, while they attract each other at
$\theta=60^\circ$ and $\theta=30^\circ$, with a stronger attraction
in the later case. This uniform attractive or repulsive behavior is
destroyed for configurations with smaller angles. In particular, in the case of
$\theta=0$, the particles attract each other in the close-contact regime, $d\lesssim2.7R$,
while they repel each other for the larger separations. We will show later
that this peculiar behavior is associated to the spontaneous broken axial symmetry
of the defect pairs.

\begin{figure}
\begin{center}
\includegraphics[width=6.25cm, angle=-90]{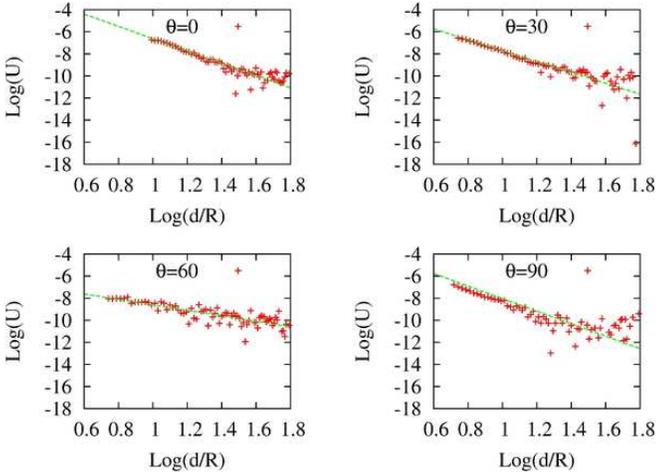}
\caption{Log-log plots of the effective potential energy vs. inter-particle
separation shows an orientation-dependent asymptotic power-law
behavior. Fitting a two-parameter function $c_{0}+c_{1}(d/R)^{c_2}$
to $U(d,\theta)$, we get $c_2 = -5.7 \pm 0.2$, $-5.0 \pm 0.2$,
$-2.5 \pm 0.4$ and $-5.7 \pm 0.2$ for $\theta=0$, $30^\circ$, $60^\circ$ and $90^\circ$ respectively.}
\label{pot-log}
\end{center}
\end{figure}

In order to analyze the distance dependence of the effective potential,
we have fitted a two-parameter function $c_{0}+c_{1}(d/R)^{c_2}$ to the plots of Fig.~\ref{pot-distance}.
The fits are shown in Fig.~\ref{pot-log} and correspond to exponents $c_2 = -5.6 \pm 0.2$
and $c_2 = -5.7 \pm 0.2$ for $\theta=0^\circ$ and $\theta=90^\circ$ respectively.
It is noticed that exponents are slightly larger in magnitude in comparison
to the weak anchoring analytical analysis~\cite{Ruhwandl.PhysRevE.55.97}, $c_2^{\mathrm{weak}} = -5$.
The exponent is $c_2 = -2.5 \pm 0.4$ for $\theta=60^\circ$, which describes
a deformation field of longer range in contrast to the weak anchoring theory.
Finally, in the case $\theta=30^\circ$, the exponent is
in agreement with the analytical prediction, $c_2 = -5.0 \pm 0.2$.

\begin{figure}
\begin{center}
\includegraphics[width=6.0cm, angle=-90]{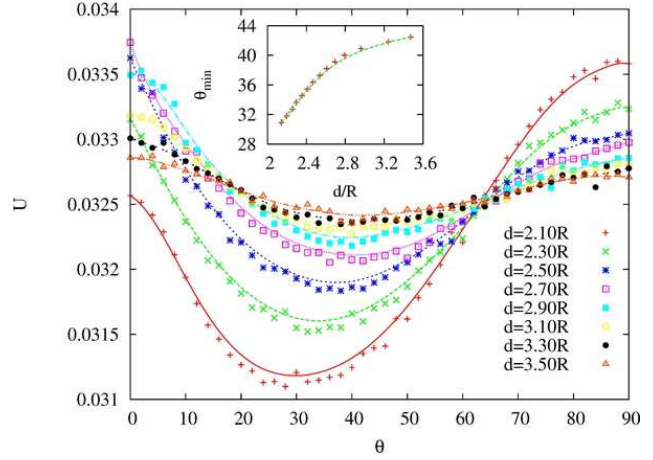}
\caption{Effective potential energy as a function of angle $\theta$
for different inter-particle separations $d/R=$ 2.10, 2.30, 2.50,
2.70, 2.90, 3.10, 3.30 and 3.50. The inset plot shows the angle at
which the free energy achieves its minimum, $\theta_{\mathrm{min}}$,
as a function of inter-particle separation. The error bars indicate
numerical uncertainty in the minimum position. The free energy
assumes its global minimum at $\theta_{\mathrm{min}} = 28^\circ \pm
2^\circ$.} \label{pot-angle}
\end{center}
\end{figure}

Fig.~\ref{pot-angle} shows the effective potential energy as a function of angle $\theta$
for fixed particle-particle separations ($d/R= 2.10$, $2.30$,
$2.50$, $2.70$, $2.90$, $3.10$, $3.30$ and $3.50$). It is noticed that
each of the plots has a unique global minimum. Moreover, the free energy minimum, as
well the orientation angle at which the minimum is achieved ($\theta_{\mathrm{min}}$),
decreases monotonously as the particles approach each other. This behavior is shown
clearly in the inset plot of Fig.~\ref{pot-angle}. Extrapolating $\theta_{\mathrm{min}}$
to the limit $d \simeq 2R$, we find that the global minimum of the free energy is
achieved for $\theta = 28^\circ \pm 2^\circ$. This result is consistent with the
experimental results of Poulin {\em et al.}~\cite{Poulin.PhysRevE.57.98}, Smaylukh {\em et al.}~\cite{Smalyukh.PhysRevLett.95.05}
and Kotar {\em et al.}~\cite{kotar.PhysRevLett.96.06}, who find the equilibrium angle
to be $\theta \simeq 30^\circ$.

\begin{figure}
\begin{center}
\includegraphics[width=6.0cm, angle=-90]{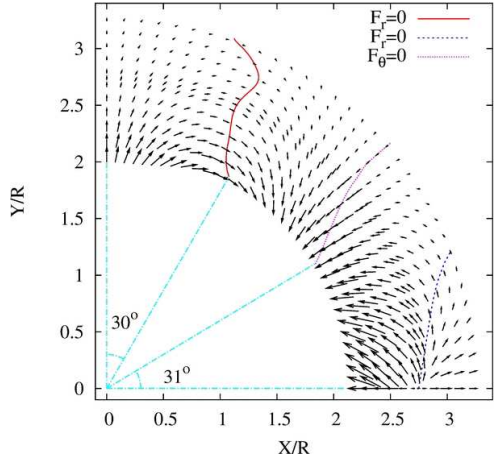}
\caption{The vector field of net force between two colloidal particles in the $x$-$y$ plane. The center of
one of the particles is fixed at the origin. $F_r$ and $F_\theta$ denote the radial and tangential
components of the force, respectively.}
\label{vector}
\end{center}
\end{figure}

The vector field of the net force between the particles can be calculated
by taking the gradient of the effective potential energy. The net force field
is shown in Fig.~\ref{vector} and the configurations at which
the radial ($F_r$) and tangential ($F_\theta$) components of the net force vanishes are indicated.
It is easily noticed that the force field drives the system towards
the configuration of minimum free energy, i.e. $\theta_{\mathrm{min}} \simeq
28^\circ$ and $d \simeq 2R$. The radial component of the net force is
positive for $\theta \gtrsim 60^\circ$ and thus, the force is repulsive. The angle
at which the radial component of net force changes sign depends on the inter-particle
separation and varies in the range $60^\circ < \theta < 70^\circ$
for the investigated configurations. Repulsive interaction is expected to show up when
$\theta > 75^\circ$ in quadrupolar approximation~\cite{Smalyukh.PhysRevLett.95.05}. We
associate this slight discrepancy to deviations from the quadrupolar approximation.

As it was mentioned earlier in this section, the inter-particle
force exhibits a non-monotonous behavior as a function of
inter-particle separation for small angles ($\theta \lesssim
15^\circ$). This behavior can be seen in
Figs.~\ref{pot-distance},~\ref{pot-contour} and~\ref{vector}. In
particular, for $\theta = 0^\circ$, the net force is repulsive for
large separations, while it becomes attractive for $\sim d/R
\lesssim 2.7$. To our knowledge, this peculiar behavior has not been
studied theoretically elsewhere.

Moreover, its experimental observation requires measurement the interaction force in separations
smaller than those reported in the experimental paper by Smalyukh {\em et al.}~\cite{Smalyukh.PhysRevLett.95.05,Smalyukh_private}.
We note that precise experimental measurement of this phenomenon can be carried out by
fixing the orientation of the colloidal particles by using line optical tweezers~\cite{Kimura_private}.

\begin{figure}
\begin{tabular}{lc}
{\large a} & \includegraphics[width=3.0cm, angle=90]{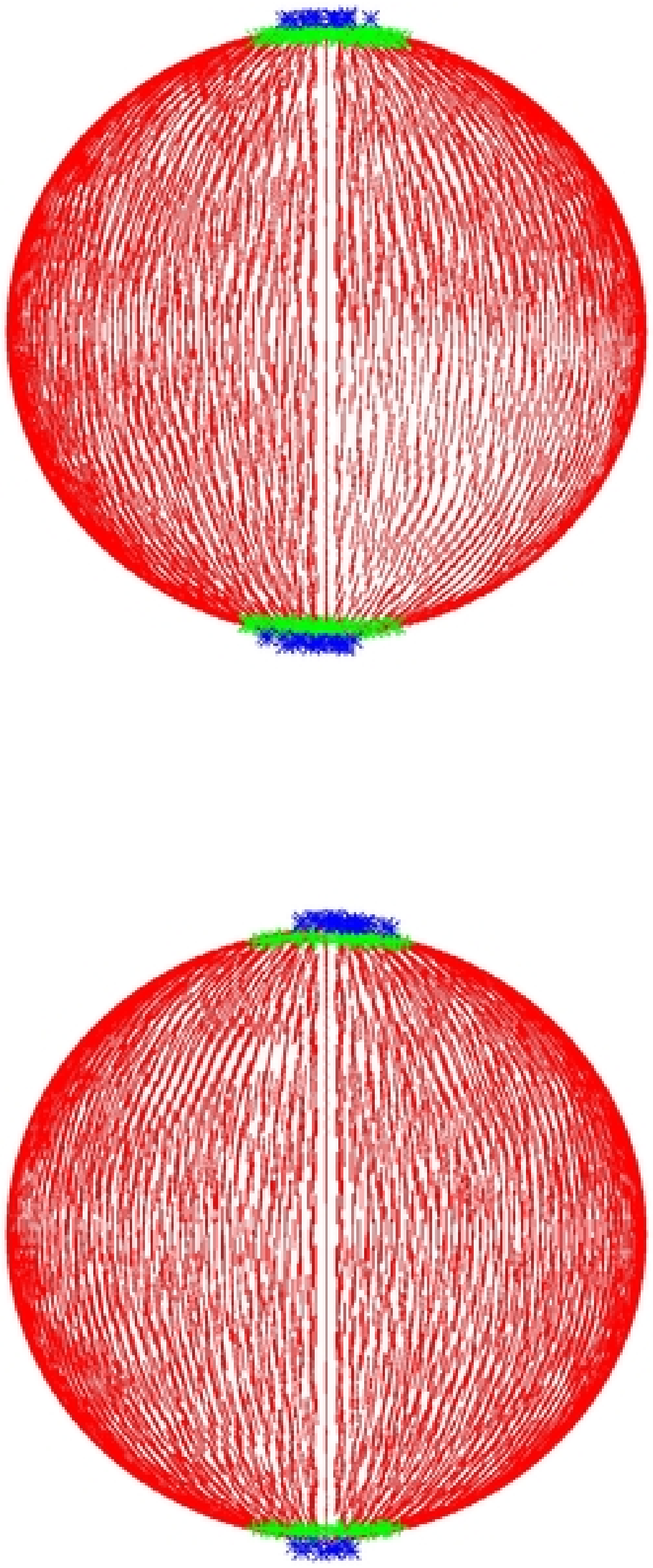} \\
{\large b} & \includegraphics[width=3.0cm, angle=90]{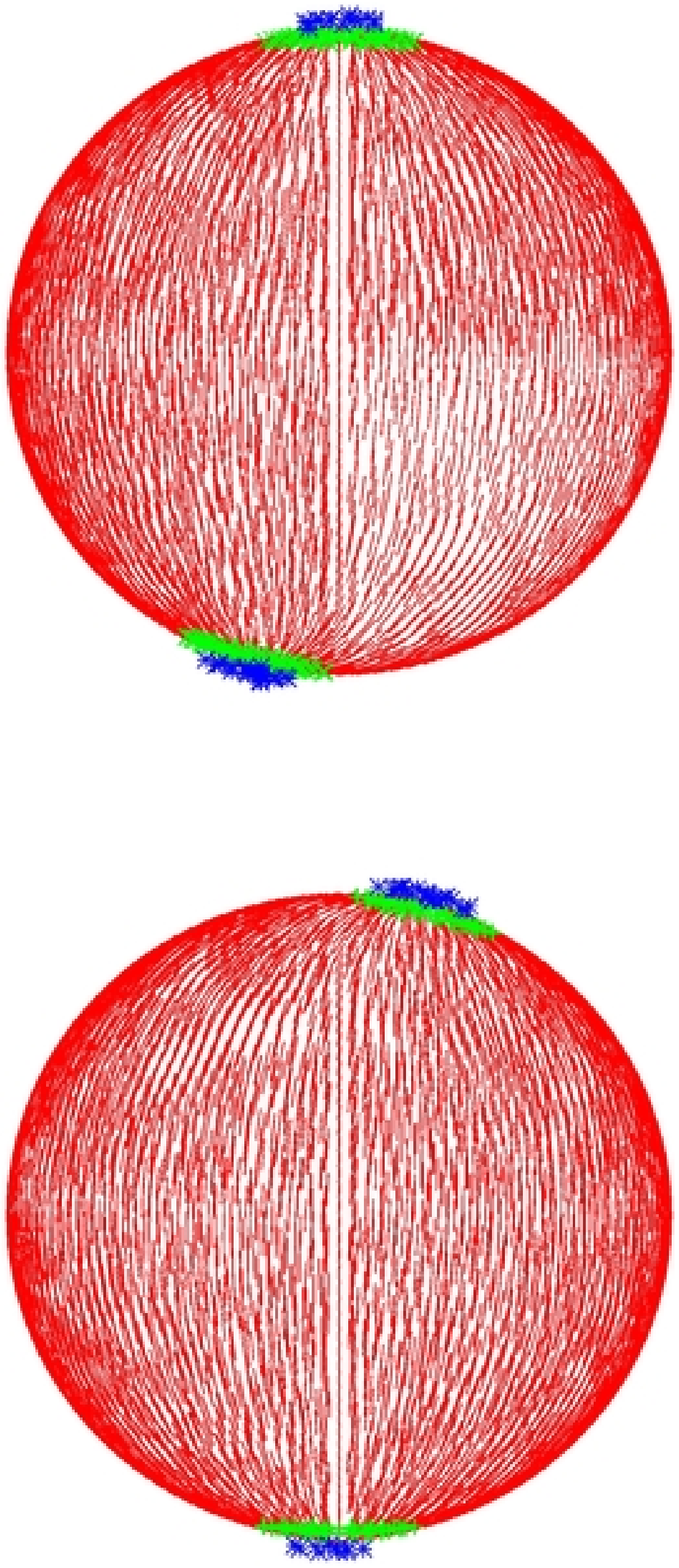} \\
{\large c} & \includegraphics[width=3.0cm, angle=90]{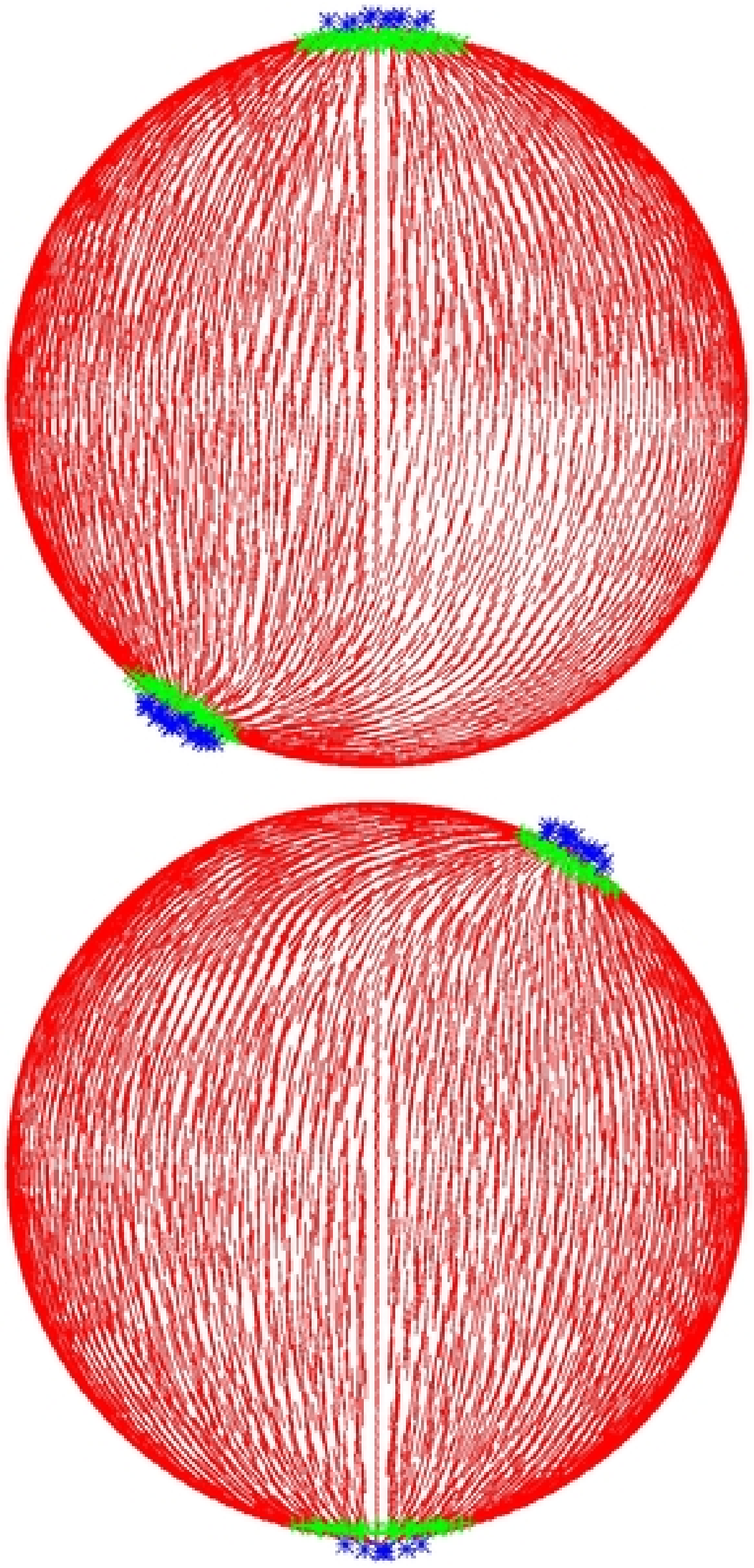} \\
& \includegraphics[width=1.0cm, angle=00]{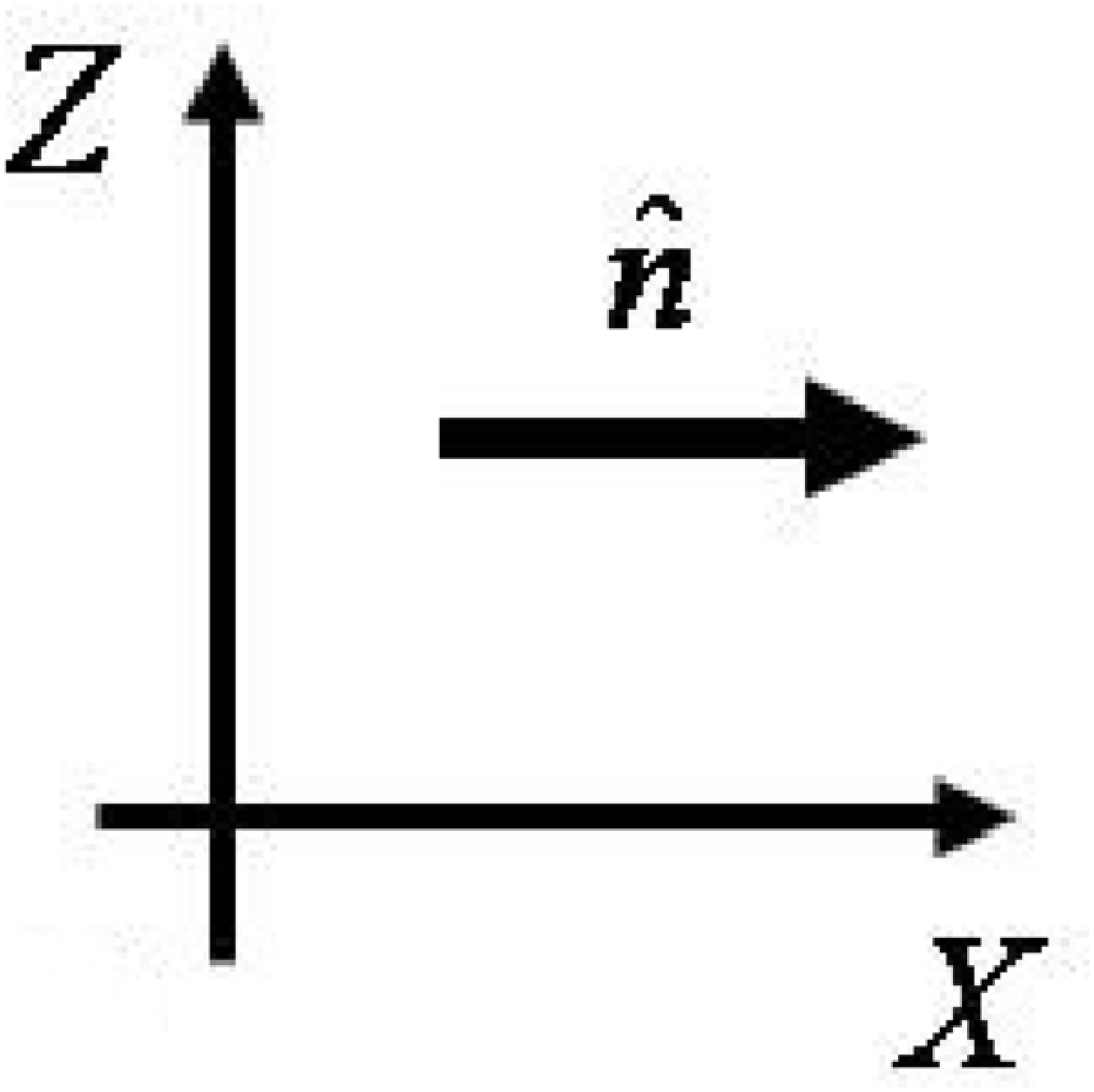}
\end{tabular}

\caption{\label{map-1} Director profile and topological defects on
the surface of the colloidal particles when they approach each other
along the nematic direction ($\theta=0$) at different inter-particle
separations (a) $d/R=3.00$, (b) $d/R=2.70$, and (c) $d/R=2.10$. The
regions where the scalar order parameter drops below $0.6 S_b$ is
highlighted, signaling the existence of a topological defect. The
defect points on the surface and bulk are indicated by blue and blue
colors, respectively.}
\end{figure}

\begin{figure}
\begin{tabular}{lc}
{\large a} & \includegraphics[width=3.0cm, angle=90]{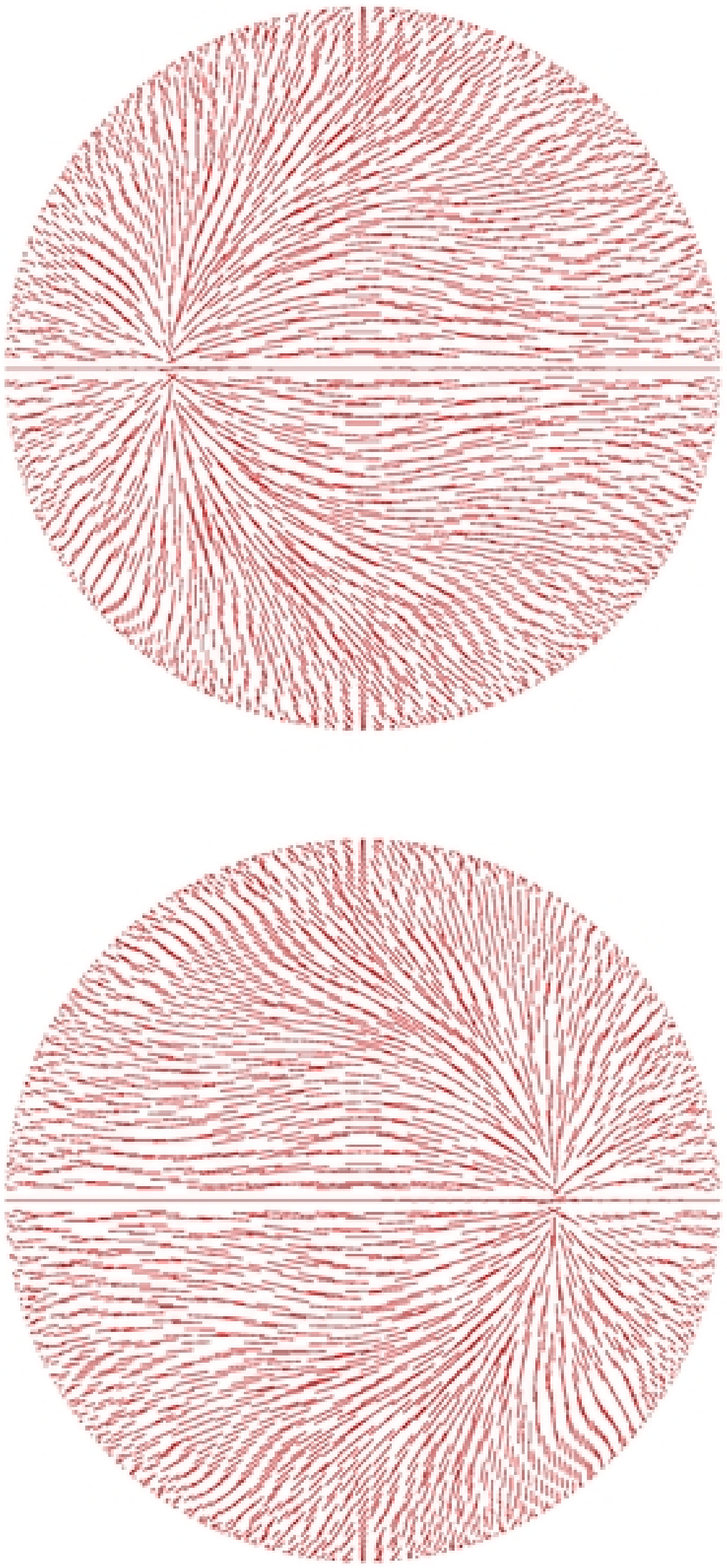} \\
{\large b} & \includegraphics[width=3.0cm, angle=90]{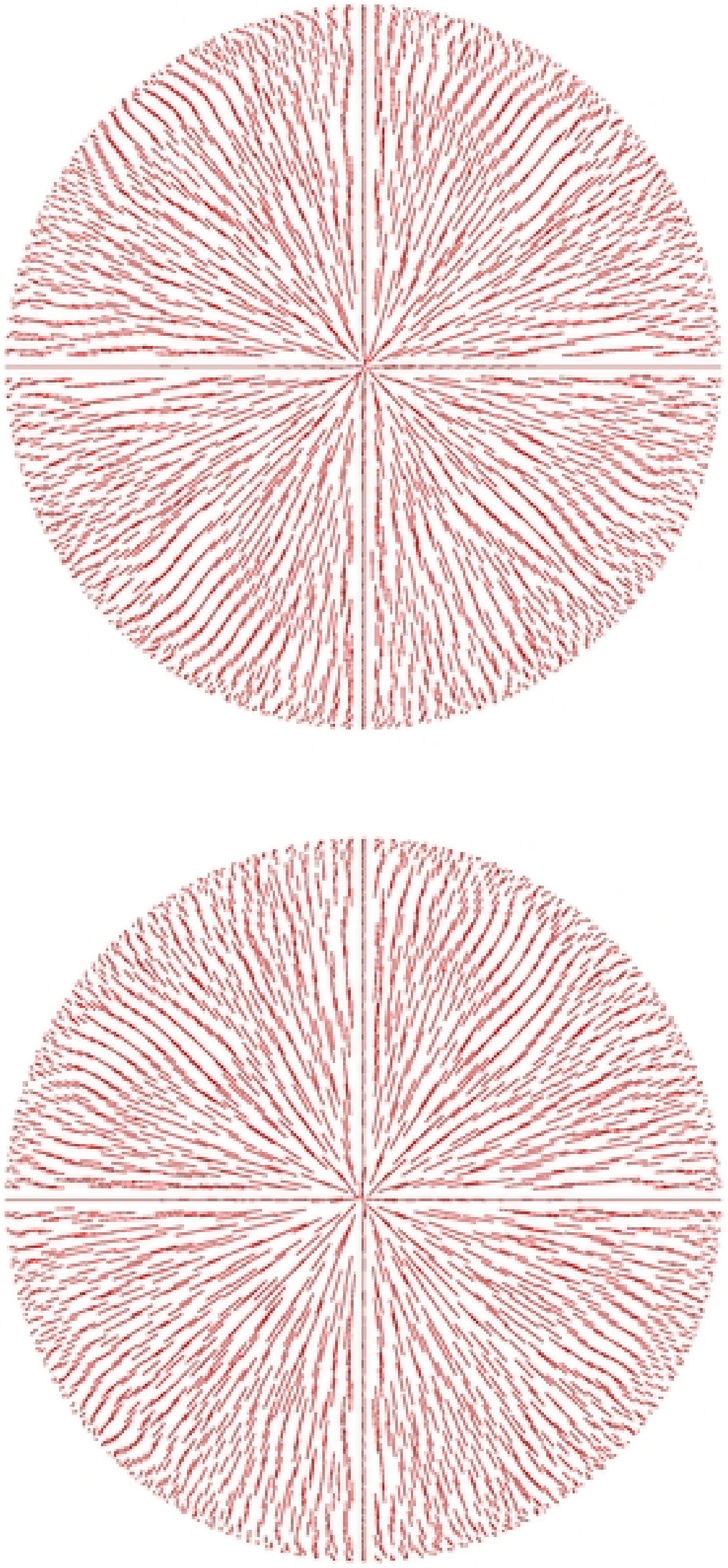} \\
& \includegraphics[width=1.0cm,angle=00]{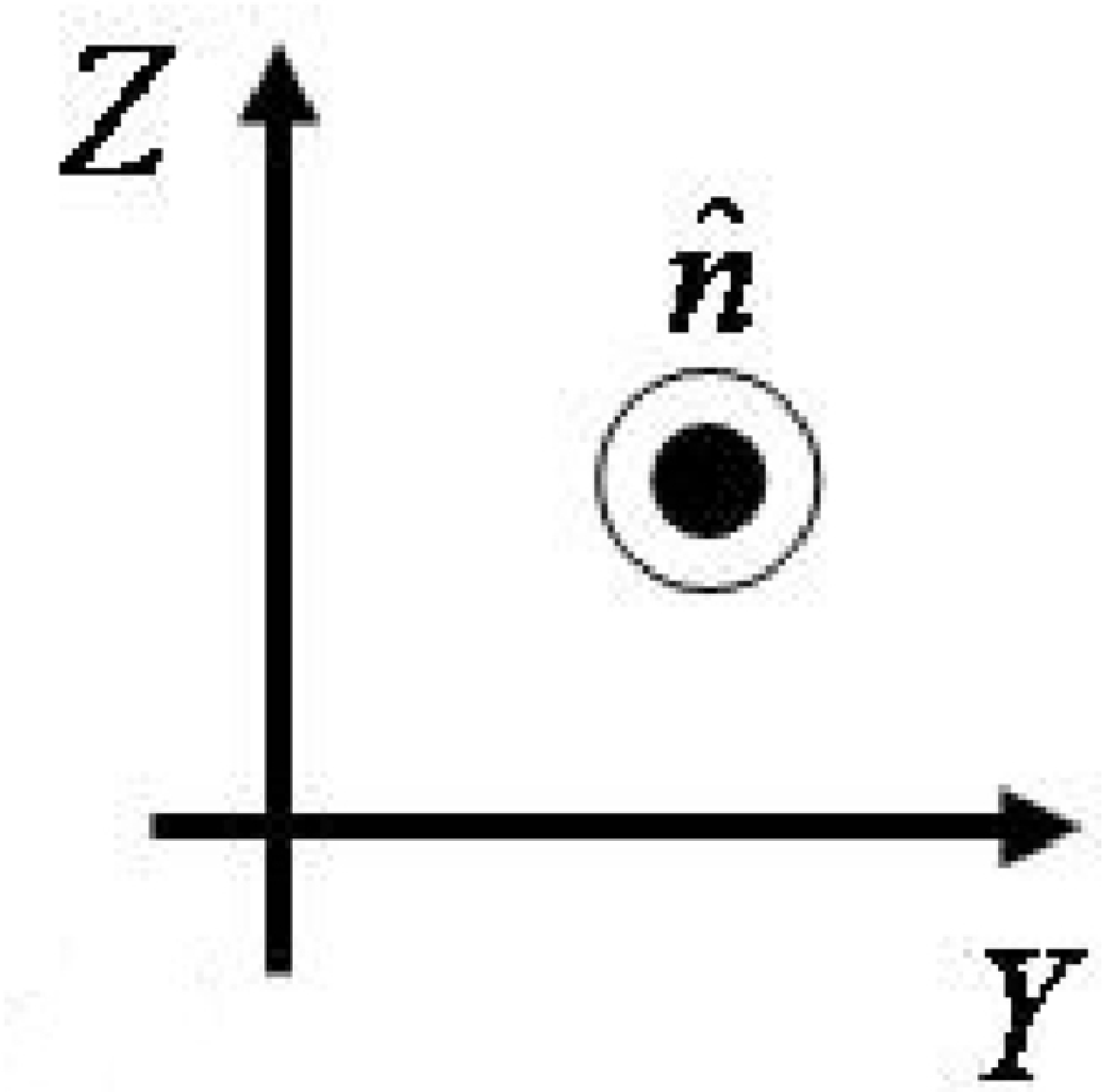}
\end{tabular}

\caption{ \label{map-2} Director profiles on the surface of the
colloids for $\theta=0$ and $d=2.10 R$, viewed along the $x$-axis.
(a) front view (near boojums) (b) back (far boojums). The
displacement of approaching boojums from the $x$-axis is clearly
noticeable.}
\end{figure}

In order to gain insight into this phenomenon, we study the configuration of the topological
defects at different inter-particle separations, as shown in Fig.~\ref{map-1}.
When the particles are well separated ($d/R \gtrsim 3$), the boojum defects are aligned on
the $x$-axis and the defect-particle pairs
have a quadrupolar symmetry for each of the colloidal particles. The repulsion
in this regime is thus associated to the head-to-head interaction of defects of
equal charge $+1/2$ (Fig.~\ref{map-1}a).  When particles approach each other, the continuous
$O(2)$ symmetry of the defect pairs is continuously broken due to the strong repulsion
between the approaching boojums, driving them away from the axis of symmetry (Fig.~\ref{map-1}b-c).
The nematic director profile on the front and back of the colloidal particles is shown in
Fig.~\ref{map-2}a and~\ref{map-2}b respectively, as seen along the $x$-axis. It is noticed that
the approaching pair of boojums are driven away from the $x$-axis (Fig.~\ref{map-2}a), while the
boojums on the back of the particles remain on the $x$-axis. The displacement of the approaching
pair of boojums induces an attraction between the colloidal particles due to the
energetic tendency to reduce the volume of the distorted region between the particles.

\begin{figure}
\begin{tabular}{lc}
{\large a} & \includegraphics[width=3.0cm, angle=90]{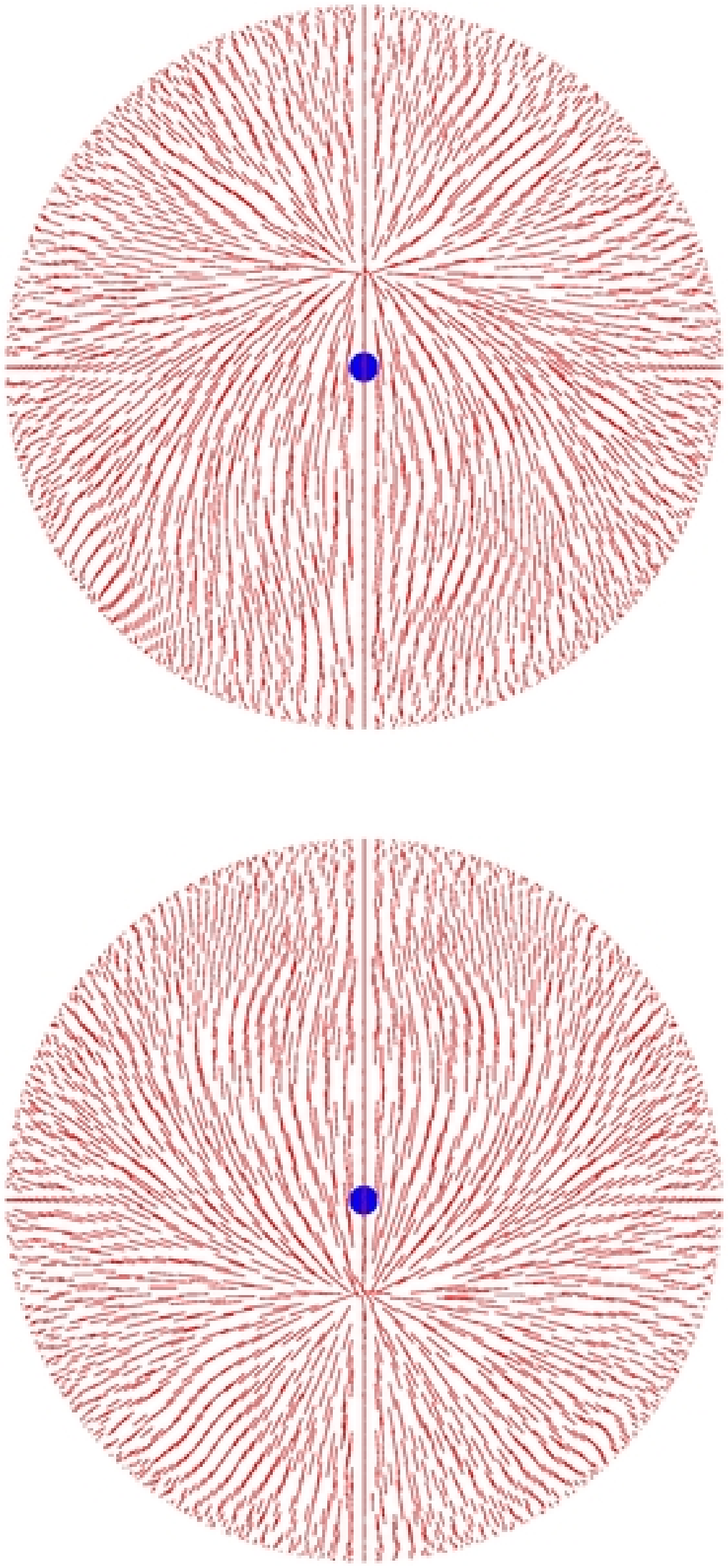} \\
{\large b} & \includegraphics[width=3.0cm, angle=90]{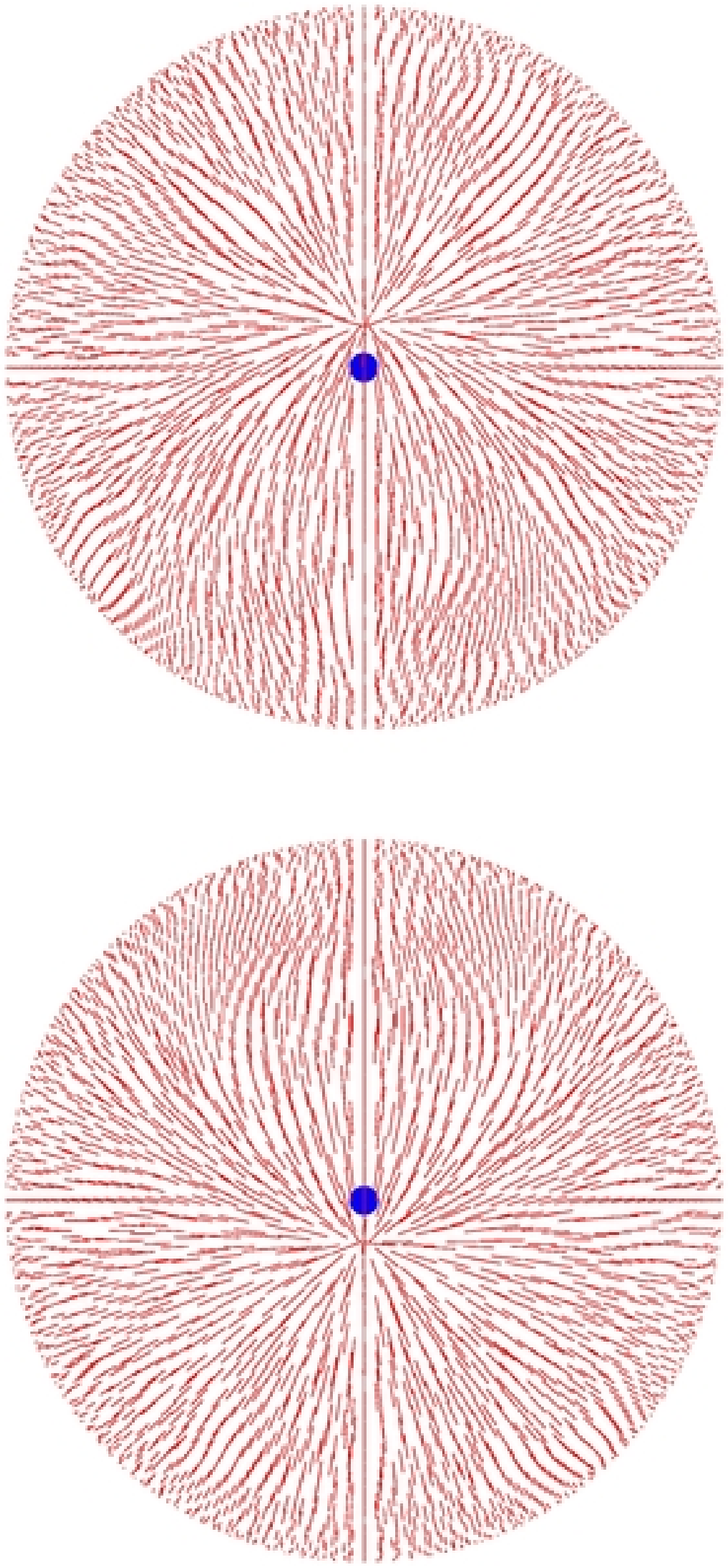} \\
{\large c} & \includegraphics[width=3.0cm, angle=90]{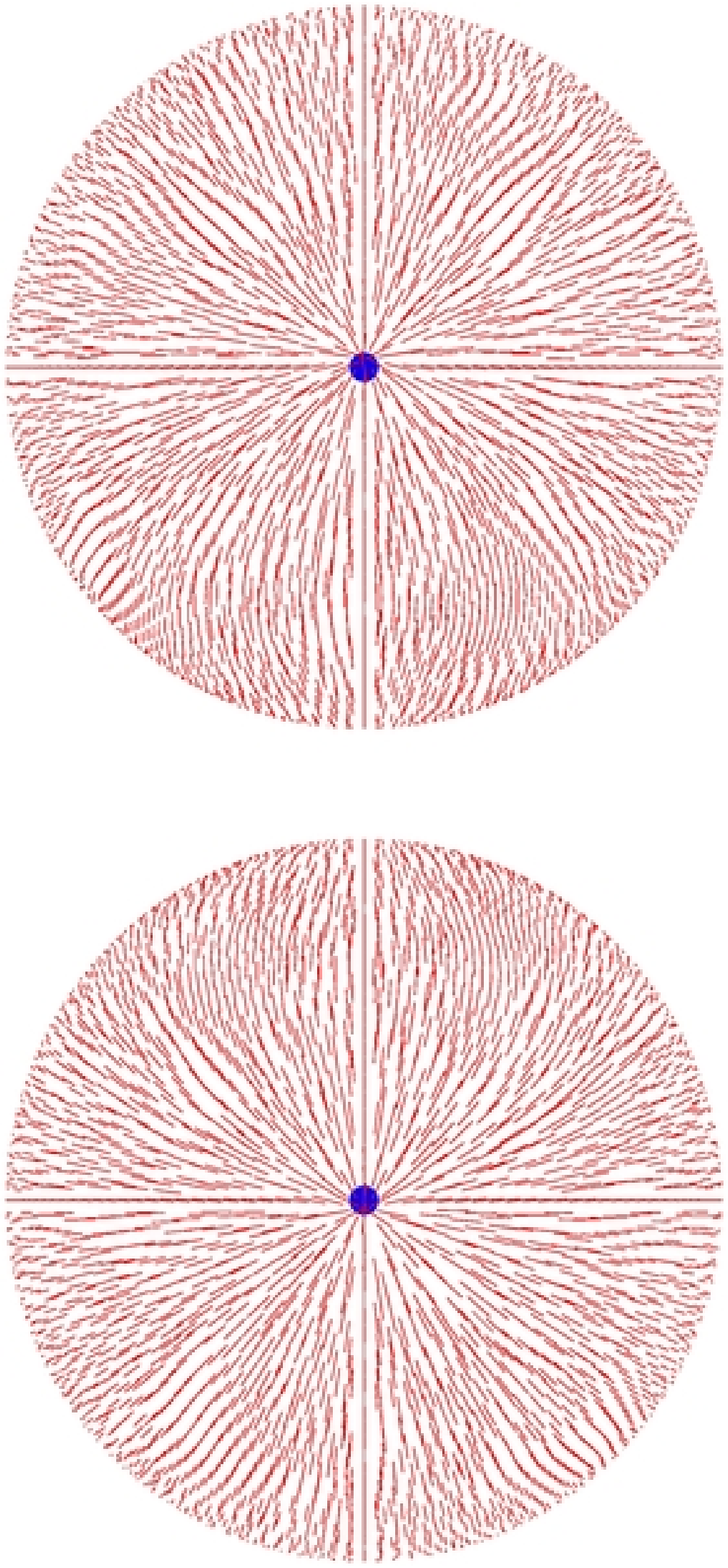} \\
{\large d} & \includegraphics[width=3.0cm, angle=90]{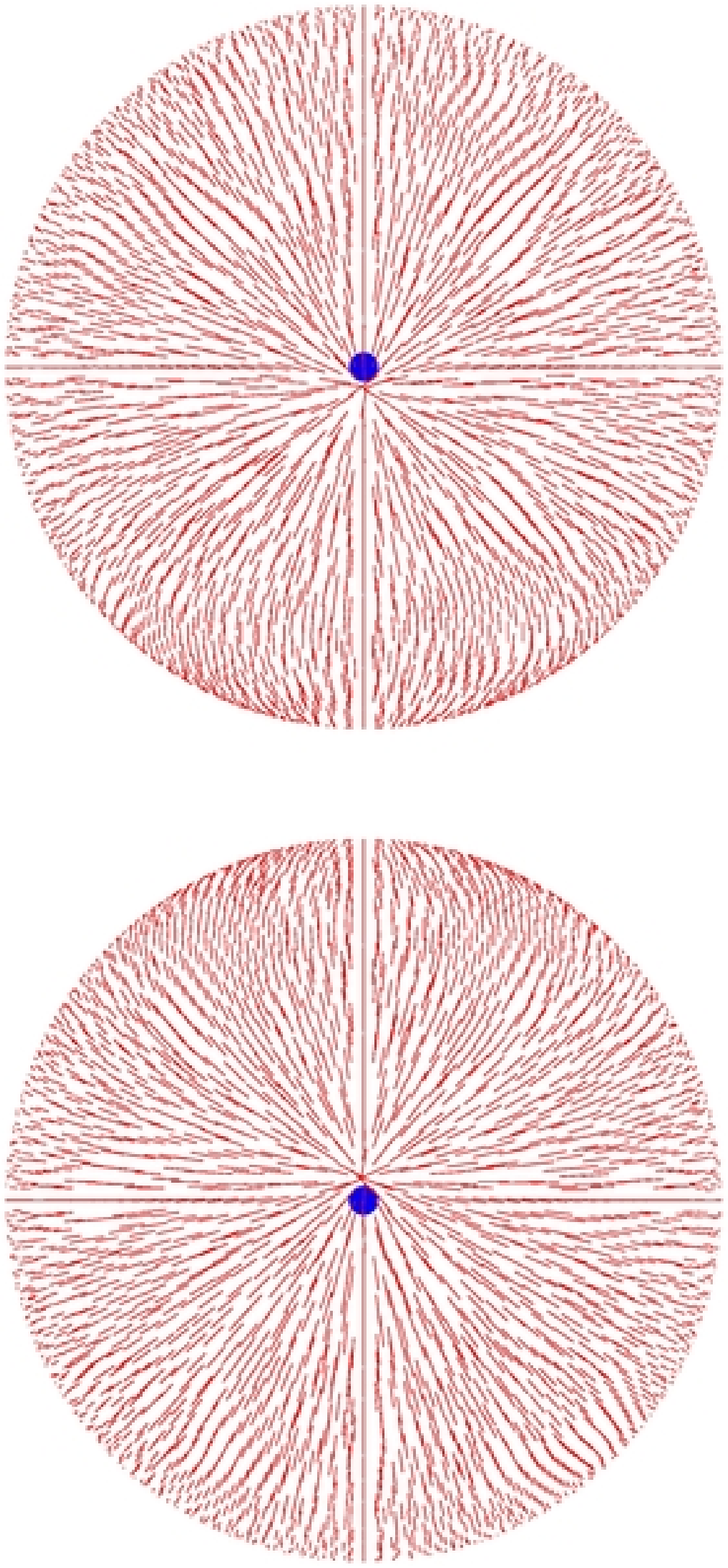} \\
& \includegraphics[width=1.0cm,angle=00]{10c.eps}
\end{tabular}

\caption{\label{map-3} Director profiles on the surface of the
colloids for a center separation of $d/R=2.10$, and for four
different orientations (a) $\theta=20^\circ$, (b) $\theta=30^\circ$,
(c) $\theta=40^\circ$ and (d) $\theta=50^\circ$. The spheres are
viewed along the bulk nematic director and from the side where the
boojums are closer to each other. The center of each of the spheres
is indicated by a blue dot. It is noticed that the axial symmetry of
the defect-pair on each of the spheres is essentially preserved.}
\end{figure}

\begin{figure}
\includegraphics[width=6.0cm, angle=270]{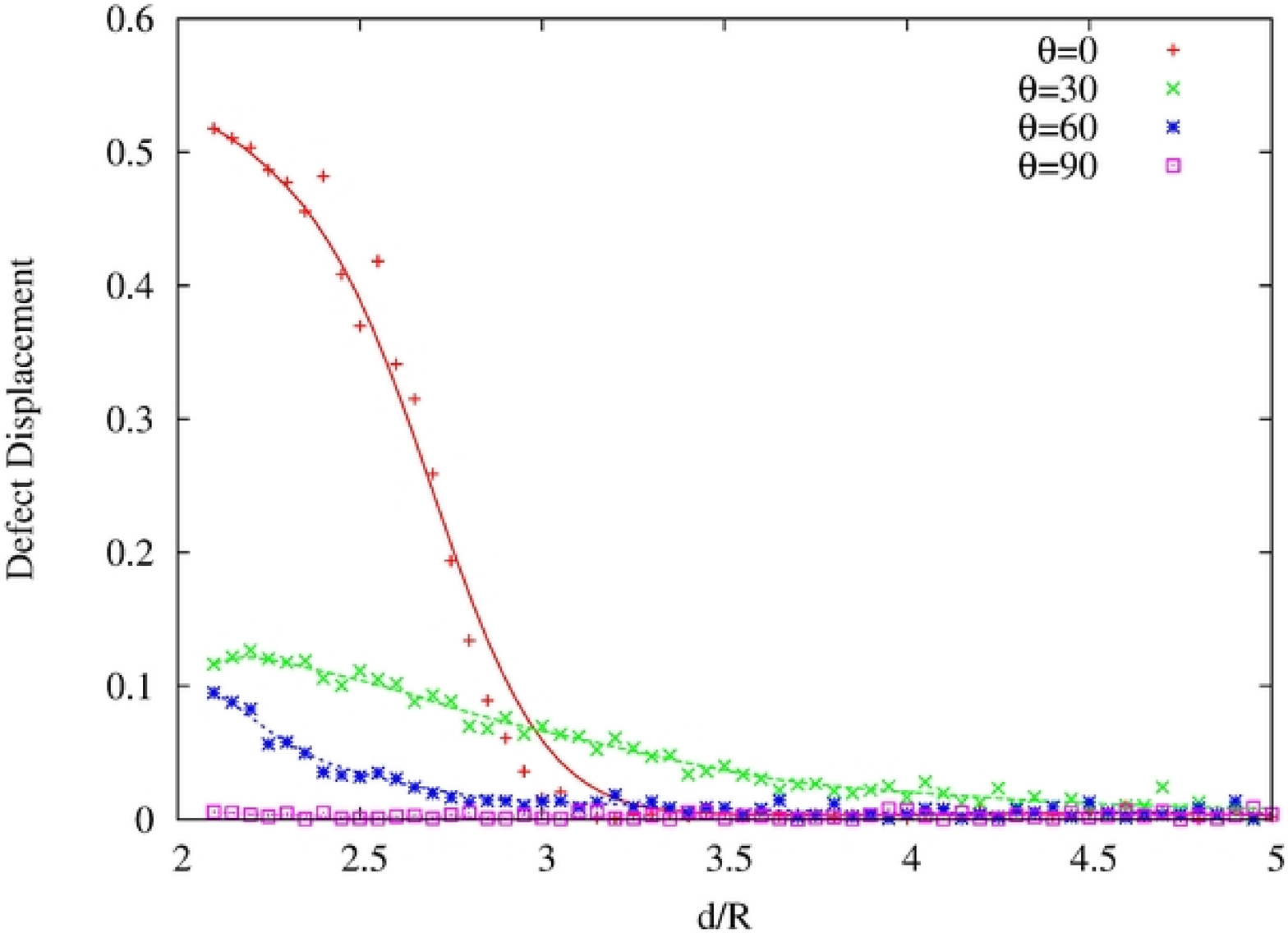}
\caption{\label{defect-displacement} Defect displacement as a
function of inter-particle distances for $\theta = 0^\circ,
30^\circ, 60^\circ$ and $90^\circ$. Displacements are scaled to the
particles radius. It shows that defects do not moves a lot for large
angles. The curves are only as eye guides.}
\end{figure}

At larger angles ($\theta \gtrsim 30^\circ$), the axial symmetry of
the defect-pair on each of the particles is almost preserved in the
whole range of inter-particle separations (see Fig.~\ref{map-3} and
Fig.~\ref{defect-displacement}). Therefore, the monotonous
attractive or repulsive interaction at larger angles can be
explained by merely taking into account the quadrupolar deformation
field of each of the particles~\cite{Ruhwandl.PhysRevE.55.97}.

We finally remark that although the calculations were carried out for a specific choice of bulk and surface
energy parameters, we expect our results to be insensitive to variations in the parameters as long
as they describe the same ($R \gg \xi$, and strong anchoring).

\section{Conclusions}\label{conc}
In this paper, we studied the interaction of two spherical colloidal particles
with degenerate planar anchoring in a nematic media by numerically minimizing
the Landau-de Gennes~\cite{deGennes.95} bulk and Fournier~\cite{Fournier.EurophysLett.72.05}
surface energy using a finite element method. Our choice of parameters belong
to the regime of large particles (in comparison to the nematic coherence length)
and strong surface anchoring. We obtained the nematic-induced effective potential
energy of the system for different inter-particle separations and orientations
with respect to the bulk nematic director.

By studying the free energy landscape of the system, we found that
the system assumes its unique global minimum of energy when the
particles are in close contact and are oriented at an angle $\theta
= 28^\circ \pm 2$ with respect to the bulk nematic director. Our
results are in a very close agreement with the experimental results
in Ref. 4, $\theta\simeq 30^\circ$. To the best of our knowledge, we
have provided the first clear theoretical explanation of these
experimental findings.

Our results suggest that for large inter-particle separations, the
quadrupolar structure of the defect-pair on each of the particles is
essentially preserved, resulting in a monotonous attractive or
repulsive inter-particle net force, depending on the orientation
angle. However, for smaller orientation angles ($\theta \gtrsim
15^\circ$) and at smaller inter-particle separations, the axial
symmetry of the defect-pairs is continuously broken, resulting in
the emergence of an attractive interaction due to the tendency of
the system to reduce the volume of distorted fluid. This very
unexpected attraction, in very short distances, has not been
reported before and may be of interest to be explored experimentally
too.

The Finite Element method, used in this study, simply can be
extended to more complicated geometries. Inter-particles interaction
for nonspherical colloids with planar
anchoring~\cite{Lapointe.science.326.09} and also many-body
interactions between the colloids in colloidal
aggregations~\cite{Tasinkevych.EurPhysJE.10.06,Araki.PhysRevLett.97.06}
are in direction of our future studies.
\\
{\bf Aknowledgement:}
We would like to thank Seyyed Mehdi Fazeli, Maryam Saadatmand, Ivan Smalyukh and Yasuyuki Kimura for their valuable comments and discussions. MRE acknowledges the Center of Excellence in Complex Systems and Condensed Matter (CSCM) for partial support.

\footnotesize{
\bibliography{mozaffari} %your .bib file

\begin{mcitethebibliography}{35}
\bibitem[Sma()]{Smalyukh_private}
Upon the completion of this work, it was brought to our attention in a private
  communication with Ivan Smalyukh that they had observed displacement of the
  topological defects in their experiments.\relax
\mciteBstWouldAddEndPunctfalse
\mciteSetBstMidEndSepPunct{\mcitedefaultmidpunct}
{}{\mcitedefaultseppunct}\relax
\EndOfBibitem
\end{mcitethebibliography}
\bibliographystyle{rsc} %the RSC's .bst file
}

\end{document}